\documentclass[aip,jcp,preprint,amsmath,amssymb]{revtex4-1} 
\usepackage{graphicx}
\usepackage{dcolumn}
\usepackage{bm}
\usepackage{amsmath}
\usepackage{amssymb}
\usepackage{txfonts}
\usepackage{url}
\usepackage{braket}
\usepackage{multirow}
\usepackage{here}
\usepackage[usenames,dvipsnames]{color}
\definecolor{Gray}{gray}{0.0}
\definecolor{lightGray}{gray}{0.35}
\newcommand{\tvb}{\textsc{TurboRVB}} 
\newcommand{\tvbg}{\textsc{Turbo-Genius}} 
\newcommand{\prep}{\textsc{prep}}

\makeatletter
\def\Hline{%
\noalign{\ifnum0=`}\fi\hrule \@height 1pt \futurelet
\reserved@a\@xhline}
\makeatother

\begin{document}
\title{Space-warp coordinate transformation for efficient ionic force calculations in quantum Monte Carlo}

\author{Kousuke Nakano}
\email{kousuke\_1123@icloud.com}
\affiliation{International School for Advanced Studies (SISSA), Via Bonomea 265, 34136, Trieste, Italy}
\affiliation{Japan Advanced Institute of Science and Technology (JAIST), Asahidai 1-1, Nomi, Ishikawa 923-1292, Japan}
\author{Abhishek Raghav}
\affiliation{Japan Advanced Institute of Science and Technology (JAIST), Asahidai 1-1, Nomi, Ishikawa 923-1292, Japan}
\author{Sandro Sorella}
\email{sorella@sissa.it}
\affiliation{International School for Advanced Studies (SISSA), Via Bonomea 265, 34136, Trieste, Italy}


\date{\today}
\begin{abstract}
Ab-initio quantum Monte Carlo (QMC) methods are a state-of-the-art computational approach to obtaining highly accurate many-body wave functions. Although QMC methods are widely used in physics and chemistry to compute ground-state energies, calculation of atomic forces is still under technical/algorithmic development. Very recently, force evaluation has started to become of paramount importance for the generation of machine-learning force-field potentials. Nevertheless, there is no consensus regarding whether an efficient algorithm is available for the QMC force evaluation, namely one that scales well with the number of electrons and the atomic numbers. In this study, we benchmark the accuracy of all-electron variational Monte Carlo (VMC) and lattice-regularized diffusion Monte Carlo (LRDMC) forces for various mono- and heteronuclear dimers ($1 \le Z \le 35$, where $Z$ is the atomic number). The VMC and LRDMC forces were calculated with and without the so-called space-warp coordinate transformation (SWCT) and appropriate regularization techniques to remove the infinite variance problem. The LRDMC forces were computed with the Reynolds (RE) and variational-drift (VD) approximations. The potential energy surfaces obtained from the LRDMC energies give equilibrium bond lengths ($r_{\rm eq}$) and harmonic frequencies ($\omega$) very close to the experimental values for all dimers, improving the corresponding VMC results. The LRDMC forces with the RE approximation improve the VMC forces, implying that it is worth computing the DMC forces beyond VMC in spite of the higher computational cost. The LRDMC forces with the VD approximations also show improvement, which unfortunately comes at a much higher computational cost in all-electron calculations. We find that the ratio of computational costs between QMC energy and forces scales as $Z^{\sim 2.5}$ without the SWCT. In contrast, the application of the SWCT makes the ratio {\it independent} of $Z$. As such, the accessible QMC system size is not affected by the evaluation of 
ionic forces but governed by the same scaling as the total energy one.
\end{abstract}
\maketitle

%
%

\section{Introduction}
%
The quantum Monte Carlo (QMC) technique{\cite{2001FOU}} is one of the state-of-the-art numerical methods for evaluating the expectation values of many-body wave functions, and it provides extremely accurate energetics. 
There are two major QMC frameworks, specifically the variational Monte Carlo (VMC) and fixed-node diffusion Monte Carlo (FNDMC) methods.{\cite{2001FOU}} 
In the VMC framework, expectation values such as energy are calculated using a given trial wave function and a Metropolis algorithm.{\cite{1993UMR, 1998STE}} Once the energy has been calculated for a given trial function, one can readily optimize the variational parameters included in the trial wave function using an appropriate method to minimize the energy and obtain the best possible representation of the ground-state wave function. One of the drawbacks of the VMC method is its strong dependence on the quality of the trial wave function.
The FNDMC framework is a stochastic method that filters out the ground state from a given trial wave function using the so-called projection technique or the power method. Since electrons are fermionic, to avoid the well-known sign-problem instability, the projection to the ground state is restricted to leave the nodal surface of the initial trial wave function unchanged, and this is usually given by the best available DFT or VMC variational guess. Despite this approximation, the FNDMC method provides results that are usually of much higher quality than the VMC results. \tvb, the QMC package used in this work, implements the lattice-regularized version of the FNDMC method,{\cite{2005CAS, 2017BEC}} which is hereafter referred to as the LRDMC method.

To date, QMC methods have been widely applied to various materials that DFT fails to cope with, such as molecular crystals,{\cite{2018ZEN,2010HON}} two-dimensional materials,{\cite{2015MOS,2018SOR,2019FRA}} superconductors,{\cite{2013CAS}} and materials at extreme pressures.{\cite{2008ATT,2015DRU,2018MAZ}} After several successful QMC applications, the field is still undergoing rapid development, and several improvements are expected from the methodological perspective. One of the major drawbacks that hinders QMC applications is the lack of a method to compute exact atomic forces. Here, ``exact'' refers to a force that is consistent with the derivative of the potential energy surface (PES), which is not always true in the QMC framework, as will be discussed later. In the VMC framework, the modern energy minimization{\cite{2007UMR, 2007SOR}} and regularization{\cite{2008ATT}} techniques provide exact force evaluations within a reasonable computational effort. In contrast, the situation is not satisfactory in the DMC framework.

Many researchers have proposed schemes for exact and approximate DMC force calculations, including Reynolds \emph{et~al.},{\cite{1986REY}} Badinski and Needs,{\cite{2008BAD2}} Assaraf and Caffarel,{\cite{2000ASS, 2011ASS}} Chiesa \emph{et~al.},{\cite{2005CHI}} Filippi \emph{et~al.},{\cite{2000FIL}} and Moroni \emph{et~al.}{\cite{2014MOR, 2021VAN}} Some of these approaches guarantee an exact force evaluation (full consistency between the PES and the forces) but become impractical as the system size increases. Other schemes are only approximately consistent but allow the computation of forces with an affordable computational cost. Therefore, to exploit QMC forces in realistic simulations, it is very important to systematically study the reliability of the approximated force evaluations. In this study, we computed the PESs and atomic forces of six dimers---H$_2$, LiH, Li$_2$, CO, MgO, and SiH---at the VMC and LRDMC levels and investigated their corresponding consistency. The LRDMC forces were computed using the Reynolds (RE) approximation{\cite{1986REY}} and the variational drift-diffusion (VD) approximation.{\cite{2014MOR}}

Another important concern in QMC calculations is the scaling of the computational cost with the system size ($N$) and the atomic number ($Z$). On the one hand, one of the advantages of the QMC framework over other accurate quantum chemistry methods is the scaling with respect to $N$. For example, the computational cost of the coupled-cluster method with single, double, and perturbative triple excitations, or CCSD(T), which is considered to be the gold standard in quantum chemistry, grows as $N^7$, while that of DMC methods grows as $N^4$ at most. On the other hand, in the QMC framework, we should consider the fact that the computational cost of QMC with respect to $Z$ is slightly worse than to $N$. The accelerated Metropolis algorithm{\cite{1993UMR2}} and the real-space double-grid method{\cite{2020NAK1}} have been devised to alleviate this problem. The scaling with $Z$ is, however, still a little higher than that with $N$; for example, it is $Z^{5.0-6.5}$ for DMC energy calculations.{\cite{1986CEP, 1987HAM, 2005MA, 2020NAK1}} Last but not least, Tiihonen \emph{et~al.}{\cite{2021TII}} showed that the scaling of QMC atomic force calculations with pseudopotentials is worse than that of energy calculations, remarkably by orders of magnitude. In detail, they claimed that, at constant computational cost and constant statistical resolution, the accessible system size scales as $Z_{\rm eff}^{-2}$, where $Z_{\rm eff}$ is the effective valence charge. This result is particularly disappointing in the QMC community since one of the most important properties in computational materials science, the atomic force, is not available for heavier atoms such as transition metals, where QMC methods have proven to be very useful in energetics.{\cite{2003WAG, 2005ALF, 2007KYL, 2007WAG, 2008KOS, 2015ZHE, 2016LUO, 2017TRA, 2017SHI, 2017ICH, 2018SHI, 2018SON, 2019ICH, 2019SHE, 2021ICH}}

In this study, we revisited and estimated the computational costs of the QMC energy and force calculations with and without the space-warp coordinate transformation (SWCT) using eight homonuclear atoms: H$_2$, Li$_2$, N$_2$, F$_2$, P$_2$, S$_2$, Cl$_2$, and Br$_2$. We found that the ratio of the computational costs between the QMC energies and forces scale as $Z^{\sim 2.5}$ without the SWCT, while the application of the SWCT makes the ratio {\it independent} of $Z$. Indeed, the accessible system size is not affected by QMC force calculations when the SWCT variance-reduction technique is applied.

%
%

\section{Implementation}
\label{implementation}
As noted above, \tvb\ implements two types of well-established QMC methods: VMC and LRDMC. In both cases, the expectation value of the energy evaluated for a given trial wave function $\Psi$ and a complementary sampling wave function $\Phi$ can be formally written as:
\begin{equation}
E = \frac{{\int {d{\mathbf{x}}{\Psi}\left( {\mathbf{x}} \right) {\Phi}\left( {\mathbf{x}} \right) \cdot \hat {\mathcal{H}} \Psi \left( {\mathbf{x}} \right)/\Psi \left( {\mathbf{x}} \right)} }}{{\int {d{\mathbf{x}}{\Psi}\left( {\mathbf{x}} \right) {\Phi}\left( {\mathbf{x}} \right)} }} = \int {d{\mathbf{x}}{E_L}\left( {\mathbf{x}} \right)\pi \left( {\mathbf{x}} \right)},
\end{equation}
where ${\mathbf{x}} = \left( {{{\mathbf{r}}_1}{\sigma _1},{{\mathbf{r}}_2}{\sigma _2}, \ldots , {{\mathbf{r}}_N}{\sigma _N}} \right)$ here and henceforth is a shorthand notation for the $N$ electron coordinates and their spins, while 
$$
{E_L}\left( {\mathbf{x}} \right) \equiv \frac{\hat {\mathcal{H}}\Psi \left( {\mathbf{x}} \right)}{\Psi \left( {\mathbf{x}} \right)} \text{~~and~~}
\pi \left( {\mathbf{x}} \right) \equiv \frac{ {\Psi}\left( {\mathbf{x}} \right) {\Phi}\left( {\mathbf{x}} \right)}{\int {d{\mathbf{x'}}{\Psi}\left( {\mathbf{x'}} \right) {\Phi}\left( {\mathbf{x'}} \right) } }
$$ 
are the so-called local energy and the probability of configuration $\mathbf{x}$, respectively. In VMC, $\Psi = \Phi$, while in DMC, $\Phi$ is the projected fixed-node ground state, which is obtained through a stochastic implementation of the power method, thus it is no longer explicitly known in a closed form.

Let ${\bf R}_{\alpha}$ be the atomic position of a nucleus $\alpha$. The atomic force acting on $\alpha$ is represented as the negative gradient of the energy with respect to ${\bf R}_{\alpha}$:
\begin{subequations}
\begin{align}
{\bf F}_{\alpha} = - \frac{dE}{d{\bf R}_{\alpha}} = &- \braket{\frac{\partial}{\partial {\bf R}_{\alpha}} E_L}_{{\Phi \Psi_{T}}} \label{eqn:hf} \\
&-\braket{(E_L - E) \left[\frac{\partial \log \Phi}{\partial {\bf R}_{\alpha}} + \frac{\partial \log \Psi}{\partial {\bf R}_{\alpha}} \right]}_{\Phi \Psi_{T}} \label{eqn:pulay} \\
&-\sum_{i=1}^{N_c}{\frac{\partial E}{\partial c_{i}}} {\frac{d c_{i}}{d {\bf R}_{\alpha}}}, \label{eqn:add}
\end{align}
\end{subequations}
wherein Eqs.~(\ref{eqn:hf}), (\ref{eqn:pulay}), and (\ref{eqn:add}) are called the Hellmann--Feynman (HF), Pulay, and variational terms, respectively. In the variational term, the set of $\{ c_1, \cdots, c_{N_{c}}\}$ are the variational parameters included in the wave function.

\subsection{Variational Monte Carlo forces}
\label{vmc}
As mentioned already, the complementary sampling wave function $\Phi$ is equal to $\Psi$ in the VMC framework. One usually ignores Eq.~(\ref{eqn:add}) when evaluating a force, resulting in
\begin{equation}
{\bf F}_{\alpha}^{\rm{VMC}} = - \braket{\frac{\partial}{\partial {\bf R}_{\alpha}} E_L}_{{\Psi_{T}^2}} - 2 \braket{(E_L - E) \left[\frac{\partial \log \Psi}{\partial {\bf R}_{\alpha}} \right]}_{\Psi_{T}^2}.
\label{eqn:vmc-force}
\end{equation}
This is the conventionally adopted expression for the VMC forces. Neglecting Eq.~(\ref{eqn:add}) is justified only when the system is at a variational minimum for all the parameters (i.e., $\frac{\partial E}{\partial c_i}=0$, $\forall i$) or when the variational parameters, which are {\it implicitly} dependent on the atomic position, do not vary with the atomic position (i.e., $\frac{d c_i}{d R_\alpha}=0$; $\forall i$), otherwise a VMC force obtained by Eq.~(\ref{eqn:vmc-force}) can be biased. Indeed, the exact force obtained from the slope of the PES is not consistent with the atomic force obtained by the above expression. This error is referred to as the ``self-consistency error.''

To obtain a statistically meaningful value of the VMC force with finite variance, the so-called ``reweighting'' technique is needed because the HF and Pulay terms may diverge when the minimum electron--nucleus distance vanishes and when an electronic configuration approaches the nodal surface, respectively. The infinite variance of the first term can be cured by the method described in Refs.~{\onlinecite{2000ASS}} and {\onlinecite{2003ASS}} or by applying the SWCT algorithm,{\cite{1989UMR, 2010SOR,2016CLA}} whereas that of the second term can be alleviated by modifying the VMC sampling distribution using a modified trial wave function that differs from the original trial wave function only in the vicinity of the nodal surface.{\cite{2008ATT}} By defining $d$ as the distance from the nodal region, the local energy ($E_L$) and the logarithmic derivative of the wave function ($\frac{\partial \log \Psi}{\partial R_{\alpha}}$) diverge as $\sim$$1/d$, leading to an infinite variance problem in Eq.~(\ref{eqn:vmc-force}). The reweighting method proposed by Attaccalite and Sorella,{\cite{2008ATT}} which we dub the AS regularization, can cancel out the divergence without introducing bias, realizing finite-variance evaluation of VMC forces.

\subsection{Diffusion Monte Carlo forces}
\label{dmc}
The situation is more complicated in the DMC framework because of the complementary sampling wave function $\Phi$. Indeed, within the DMC framework, there is no straightforward way to evaluate the logarithmic derivative of the complementary sampling wave function ($\frac{\partial \log \Phi}{\partial {\bf R}_{\alpha}}$). Therefore, to overcome this difficulty, Reynolds \emph{et~al.}{\cite{1986REY}} proposed the approximation $\frac{\partial \log \Phi}{\partial {\bf R}_{\alpha}} \approx \frac{\partial \log \Psi}{\partial {\bf R}_{\alpha}}$, leading to:
\begin{equation}
{\bf F}_{\alpha}^{{\rm{DMC}}, {\rm{RE}}} = - \braket{\frac{\partial}{\partial {\bf R}_{\alpha}} E_L}_{{\Psi_{T}\Phi}} - 2 \braket{(E_L - E) \left[\frac{\partial \log \Psi}{\partial {\bf R}_{\alpha}} \right]}_{\Psi_{T}\Phi}.
\label{eqn:dmc-Reynolds-force}
\end{equation}
The so-called ``hybrid estimator'' technique can be further used to correct the bias in the Reynolds force due to the above approximation{\cite{2003ASS}}:
\begin{equation}
{\bf F}_{\alpha}^{{\rm hybrid}} = 2 \braket{{\bf F}_{\alpha}^{{\rm DMC, RE}}}_{\Phi \Psi_{T}} - \braket{{\bf F}_{\alpha}^{{\rm VMC}}}_{\Psi_{T}^2},
\label{eqn:dmc-hybrid-force}
\end{equation}
which is dubbed the ``hybrid force.''

Later on, a more sophisticated approximation, the VD approximation, was proposed by Filippi and Umrigar,{\cite{2000FIL}} and this was recently revisited by Moroni \emph{et~al.}{\cite{2014MOR}} This approximation replaces the logarithmic derivative of the drift-diffusion part of the Green function along the random walk with the logarithmic derivative of the trial wave function at the current configuration and the sum of the derivative of the local energy along with the previous $k$ steps. Indeed, the VD force is evaluated as:
\begin{equation}
{\bf F}_{\alpha}^{{\rm{DMC}}, {\rm{VD}}} = - \braket{\frac{\partial}{\partial {\bf R}_{\alpha}} E_L\left({\bf{x}}_{n}\right)}_{{\Psi_{T}\Phi}} - \braket{(E_L\left({\bf{x}}_{n}\right) - E) \left[2 \frac{\partial \log \Psi}{\partial {\bf R}_{\alpha}} - \frac{1}{2}\tau \sum_{n-k}^{n-1} \frac{\partial}{\partial {\bf R}_{\alpha}}[E_L\left({\bf{x}}_{i+1}\right) + E_L\left({\bf{x}}_{i}\right)] \right]}_{\Psi_{T}\Phi},
\label{eqn:dmc-VD-force}
\end{equation}
where $\tau$ is the time step defining the discretized times $t_n= n\tau$ when the local energy $E_L\left({\bf{x}}_{n}\right)$ is evaluated. This formulation was originally proposed within the standard DMC framework,{\cite{2014MOR}} but it can also be used in the LRDMC framework. Notice also that, due to time discretization, a bias proportional to $O(\tau^2)$ is also present within LRDMC, which otherwise has only finite lattice mesh error. We have chosen in all forthcoming calculations a time step $\tau=0.1$\,a.u., small enough to have negligible errors of this type. Moreover, $k$ is a hyperparameter, whose convergence was checked for each dimer.

The AS regularization has also been used for DMC force calculations,{\cite{2010VAD, 2014MOR}} but it is not an optimal regularization for this purpose because it enforces a finite density of walkers on the nodal surface.{\cite{2021VAN}} In this study, we employed the novel regularization technique recently proposed by Pathak and Wagner{\cite{2020PAT}} for both the RE and VD forces, which we shall refer to as the PW regularization. The PW regularization was originally proposed for computing the derivative of the wave function in the VMC optimization,{\cite{2020PAT}} but it is also applicable to DMC force calculations, as proposed in Ref.~{\onlinecite{2021VAN}}. The PW regularization multiplies the terms in the brackets of Eqs.~(\ref{eqn:dmc-Reynolds-force}) and~(\ref{eqn:dmc-VD-force}) by the polynomial reweighting function $f_{\epsilon} = 7|d/\epsilon|^6 - 15|d/\epsilon|^4 + 9|d/\epsilon|^2$ when $|d/\epsilon| < 1$, where $d$ is the distance from the nodal surface, defined by $ d= \Psi({\vec x})/ \lVert\nabla \Psi({\vec x}) \rVert $. 

\subsection{Space-warp coordinate transformation}
We employed the SWCT{~\cite{1989UMR, 2010SOR}} in both the VMC and LRDMC force calculations. The SWCT is used to mimic the displacement of charges around the nucleus,
\begin{equation}
{{\mathbf{\bar r}}_i} = {{\mathbf{r}}_i} + \Delta {{\mathbf{R}}_{\alpha}}{\omega _{\alpha}}\left( {{{\mathbf{r}}_i}} \right), 
\end{equation}
\begin{equation}
{\omega _{\alpha}}\left( {\mathbf{r}} \right) = \frac{{ \kappa \left( {\left| {{\mathbf{r}} - {{\mathbf{R}}_{\alpha}}} \right|} \right)}}{{\sum\nolimits_{\beta = 1}^M {\kappa\left( {\left| {{\mathbf{r}} - {{\mathbf{R}}_{\beta}}} \right|} \right)} }},
\end{equation}
where ${\kappa\left( r \right)}$ is a function that decays sufficiently fast for large $r$ because the charges far from the nuclei should not be affected by the SWCT, and $\omega\to 0$ as a consequence of this requirement. It turns out that the original\cite{1989UMR} simple choice $\kappa\left( r \right) = 1/{r^4}$ works very well, and this is indeed adopted in \tvb. The atomic force is evaluated with the SWCT{~\cite{2010SOR}}:
\begin{equation}
\begin{split}
{{\mathbf{F}}_{\alpha}} = &- \braket{
\frac{\partial }{{\partial {{\mathbf{R}}_{\alpha}}}}{E_L} + \sum\limits_i^{} {{\omega _{\alpha}}\left( {{{\mathbf{r}}_i}} \right)\frac{\partial }{{\partial {{\mathbf{r}}_i}}}{E_L}}
}_{{\Psi_{T}\Phi}} \\
&- 2 \braket{ ({{E_L}} - E) \left\{ \frac{\partial }{{\partial {{\mathbf{R}}_{\alpha}}}}\log \left( \Psi \right) + \sum\limits_i^{} {\left[ {{\omega _{\alpha}}\left( {{{\mathbf{r}}_i}} \right)\frac{\partial }{{\partial {{\mathbf{r}}_i}}}\log \left( \Psi \right) + \frac{1}{2}\frac{\partial }{{\partial {{\mathbf{r}}_i}}}{\omega _{\alpha}}\left( {{{\mathbf{r}}_i}} \right)} \right] } \right\}}_{{\Psi_{T}\Phi}}
\label{forces:warp-RE}
\end{split}
\end{equation}
for the VMC ($\Phi = \Psi_{T}$) and RE-LRDMC forces, and
\begin{equation}
\begin{split}
{{\mathbf{F}}_{\alpha}} & = {\rm Eq.}~(\ref{forces:warp-RE}) \\
& + \braket{\frac{E_L\left({\bf{x}}_{n}\right)-E}{2}\tau \sum_{j=n-k}^{n-1} \left\{ \frac{\partial}{\partial {\bf R}_{\alpha}}[E_L\left({\bf{x}}_{j+1}\right) + E_L\left({\bf{x}}_{j}\right)] + \sum\limits_i^{} {\omega _{\alpha}}\left( {{{\mathbf{r}}_i^{j+1}}} \right)\frac{\partial }{{\partial {{\mathbf{r}}_i^{j+1}}}} E_L\left({\bf{x}}_{j+1}\right) +{\omega _{\alpha}}\left( {{{\mathbf{r}}_i^{j}}} \right)\frac{\partial }{{\partial {{\mathbf{r}}_i^{j}}}} E_L\left({\bf{x}}_{j}\right) \right\}}_{\Psi_{T}\Phi}
\end{split}
\label{forces:warp-VD}
\end{equation}
for the VD-LRDMC force, where the brackets indicate a Monte Carlo average over the trial wave function and $\mathbf{r}_i^{j}$ are the electronic coordinates corresponding to ${\bf{x}}_{j}$. All the terms above can be very efficiently evaluated by the adjoint algorithmic differentiation in \tvb.{~\cite{2010SOR}}

%
%

\section{Computational details}
\label{computational_details}
The all-electron VMC and LRDMC calculations for the selected dimers (H$_2$, LiH, Li$_2$, N$_2$, CO, F$_2$, MgO, SiH, P$_2$, S$_2$, Cl$_2$, and Br$_2$) were performed using \tvb,{\cite{2020NAK2}} and this was combined with the Python package \tvbg{\cite{2020NAK2}} to avoid any human error and to achieve efficient calculations. \tvb\ employs the Jastrow antisymmetrized geminal power (JAGP){~\cite{2003CAS}} ansatz. The ansatz is composed of a Jastrow and an antisymmetric part ($\Psi = J \cdot {\Phi _{{\text{AGP}}}}$). The singlet antisymmetric part is denoted as the singlet antisymmetrized geminal power (AGPs), which reads:
\begin{equation}
{\Psi _{{\text{AGPs}}}}\left( {{{\mathbf{r}}_1}, \ldots ,{{\mathbf{r}}_N}} \right) = {\hat A} \left[ {\Phi \left( {{\mathbf{r}}_1^ \uparrow ,{\mathbf{r}}_1^ \downarrow } \right)\Phi \left( {{\mathbf{r}}_2^ \uparrow ,{\mathbf{r}}_2^ \downarrow } \right) \cdots \Phi \left( {{\mathbf{r}}_{N/2}^ \uparrow ,{\mathbf{r}}_{N/2}^ \downarrow } \right)} \right],
\end{equation}
where ${\hat A}$ is the antisymmetrization operator, and $\Phi \left( {{\mathbf{r}}_{}^ \uparrow ,{\mathbf{r}}_{}^ \downarrow } \right)$ is called the pairing function. The spatial part of the geminal function is expanded over the Gaussian-type atomic orbitals:
\begin{equation}
{\Phi _{{\text{AGPs}}}}\left( {{{\mathbf{r}}_i},{{\mathbf{r}}_j}} \right) = \sum\limits_{l,m,a,b} {{f_{\left\{ {a,l} \right\},\left\{ {b,m} \right\}}}{\psi _{a,l}}\left( {{{\mathbf{r}}_i}} \right){\psi _{b,m}}\left( {{{\mathbf{r}}_j}} \right)},
\label{agp_expansion}
\end{equation}
where ${\psi _{a,l}}$ and ${\psi _{b,m}}$ are primitive Gaussian atomic orbitals, their indices $l$ and $m$ indicate different orbitals centered on atoms $a$ and $b$, and $i$ and $j$ are coordinates of spin-up and spin-down electrons, respectively. When the JAGPs is expanded over $p$ molecular orbitals, where $p$ is equal to half the total number of electrons ($N_{\rm el}/2$), the JAGPs coincides with the Jastrow--Slater determinant (JSD) ansatz.{\cite{2017BEC, 2009MAR}} The JSD with DFT orbitals is referred to as JDFT hereafter. Indeed, the coefficients in the pairing functions (i.e., variational parameters in the AS part) are obtained by the built-in DFT code, named \prep, and the coefficients are fixed during the VMC optimization step. This is the simplest choice, but it is reasonable in the majority of QMC applications. The coefficients are further variationally optimized with the JSD and JAGPs ansatz at the VMC level, leading to better nodal surfaces and variational energies at the DMC level.{\cite{2019NAK}} In this study, we employed the JDFT ansatz except for in the case of the Li$_2$ dimer. The JDFT, JSD, and JAGPs ansatz were employed for the Li$_2$ dimer to further investigate the effect of orbital optimization on the VMC and LRDMC forces. The Jastrow term is composed of one-body, two-body, and three-/four-body factors ($J = {J_1}{J_2}{J_{3/4}}$).The one-body and two-body factors are used to fulfill the electron--ion and electron--electron cusp conditions, respectively, and the three-/four-body factor is employed to consider the further electron--electron correlation. The one-body Jastrow factor reads:
\begin{equation}
{J_1}\left( {{{\mathbf{r}}_1}, \ldots , {{\mathbf{r}}_N}} \right) = \exp \left( {\sum\limits_{i,I,l} {g_{I,l}^{}\chi _{I,l}^{}\left( {{{\mathbf{r}}_i}} \right)} } \right) \cdot \prod\limits_i {{{\tilde J}_1}\left( {{{\mathbf{r}}_i}} \right)},
\label{onebody_jas}
\end{equation}
\begin{equation}
{\tilde J_1}\left( {\mathbf{r}} \right) = \exp \left( {\sum\limits_I { - {{\left( {2{Z_I}} \right)}^{3/4}}u\left( {2{Z_I}^{1/4}\left| {{\mathbf{r}} - {{\mathbf{R}}_I}} \right|} \right)} } \right),
\label{onebody_j_single}
\end{equation}
where ${{{\mathbf{r}}_i}}$ are the electron positions, ${{{\mathbf{R}}_I}}$ are the atomic positions with corresponding atomic number $Z_I$, $l$ runs over atomic orbitals ${\chi _{I,l}^J}$ centered on atom $I$, and ${u\left( r \right)}$ contains a variational parameter $b$:
\begin{equation}
u\left( r \right) = \frac{b}{2}\left( {1 - {e^{ - r/b}}} \right).
\label{onebody_u}
\end{equation}
The two-body Jastrow factor is defined as:
\begin{equation}
{J_2}\left( {{{\mathbf{r}}_1}, \ldots , {{\mathbf{r}}_N}} \right) = \exp \left( {\sum\limits_{i < j} {v\left( {\left| {{{\mathbf{r}}_i} - {{\mathbf{r}}_j}} \right|} \right)} } \right),
\label{twobody_jastrow}
\end{equation}
where $v\left( r \right)$ is:
\begin{equation}
v\left( r \right) = \frac{1}{2}r \cdot {\left( {1 - F \cdot r} \right)^{ - 1}}
\label{twobody_v}
\end{equation}
and $F$ is a variational parameter. The three-body Jastrow factor is:
\begin{equation}
{J_{3/4}}\left( {{{\mathbf{r}}_1}, \ldots , {{\mathbf{r}}_N}} \right) = \exp \left( {\sum\limits_{i < j} {{\Phi _{{\text{Jas}}}}\left( {{{\mathbf{r}}_i},{{\mathbf{r}}_j}} \right)} } \right),
\end{equation}
and
\begin{equation}
{\Phi _{{\text{Jas}}}}\left( {{{\mathbf{r}}_i},{{\mathbf{r}}_j}} \right) = \sum\limits_{l,m,a,b} {g_{a,l,m,b}^{}\chi _{a,l}^{{\text{Jas}}}\left( {{{\mathbf{r}}_i}} \right)\chi _{b,m}^{{\text{Jas}}}\left( {{{\mathbf{r}}_j}} \right)},
\label{threebody_jas}
\end{equation}
where the indices $l$ and $m$ again indicate different orbitals centered on corresponding atoms $a$ and $b$. Only three-body ($a = b$) factors were used in this study.

We employed {\it modified} cc-pVTZ basis sets taken from the EMSL Basis Set Library{~\cite{2019PRI}} for the AS parts. Here, ``modified'' refers to the modification of the original basis set such that the $s$ orbitals whose exponents are larger than $8 \cdot Z^2$ (where $Z$ is the atomic number) are disregarded, and they are implicitly compensated by the homogeneous one-body Jastrow part [${\tilde J_1}\left( {\mathbf{r}} \right)$] to fulfill the electron--ion cusp condition explicitly, even at the DFT level.{\cite{2019NAK}} Indeed, a single-particle orbital is modified as:
\begin{equation}
\tilde \phi _j^b\left( {{\mathbf{r}} - {{\mathbf{R}}_b}} \right) = \phi _j^b\left( {{\mathbf{r}} - {{\mathbf{R}}_b}} \right){{\tilde J}_1}\left( {\mathbf{r}} \right),
\label{onebody_j_single_DFT}
\end{equation}
where ${{\tilde J}_1}\left( {\mathbf{r}} \right)$ is the same as in Eq.~(\ref{onebody_j_single}), and the parameter $b$ in Eq.~(\ref{onebody_u}) is optimized by direct minimization of the chosen DFT energy functional. In this way, each element of the modified basis set satisfies the so-called electron--ion cusp conditions. For the inhomogeneous and three-body Jastrow parts, we employed the cc-pVDZ basis set taken from the EMSL Basis Set Library{\cite{2019PRI}} and neglected the large exponents according to the same criteria as in the AS part. The variational parameters in the AS and Jastrow parts were optimized by the modified linear method~\cite{2007UMR, 2020NAK2} implemented in \tvb. The exponents of the Jastrow and determinant basis sets were also optimized at the VMC level with the JSD and JAGPs ansatz. The LRDMC calculations for computing the PESs were performed by the single-grid scheme~\cite{2005CAS} with lattice spaces $a$ = 0.30, 0.10, 0.10, 0.05, 0.05, and 0.03 Bohr for H$_2$, LiH, Li$_2$, CO, MgO, and SiH, respectively. We chose 0.001, 0.001, 0.001, 0.01, 0.02, and 0.01 Bohr for H$_2$, LiH, Li$_2$, CO, MgO, and SiH, respectively, for the PW regularization parameter ($\epsilon$). These are compatible with the values proposed in the original paper{\cite{2020PAT}} ($\sim$$10^{-2}$). It turned out that the choice of $\epsilon$ did not introduce bias (less than 0.5\% error in the DMC force evaluations). The LRDMC calculations for investigating the computational costs of the QMC energies and forces were also performed by the single-grid scheme~\cite{2005CAS} with lattice spaces, $a = 1/(2.0 \cdot Z)$ Bohr for H$_2$, Li$_2$, N$_2$, F$_2$, P$_2$, S$_2$, Cl$_2$, and Br$_2$. The regularization parameter $\epsilon$ was set to 0.001 (Bohr) for all the dimers.

\begin{figure*}[htbp]
  \centering
  \includegraphics[width=16.4cm]{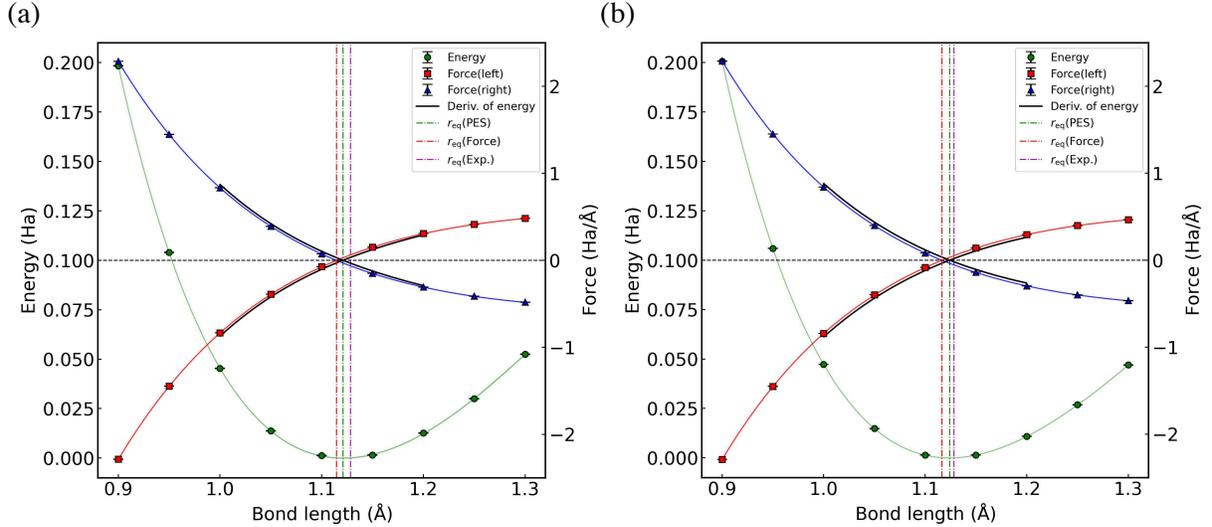}
  \caption{Potential energy curves of the CO dimer obtained by (a)~VMC calculations with the AS regularization and (b)~LRDMC calculations with the PW regularization. The JDFT ansatz was employed in these calculations. The derivatives and atomic forces acting on the left (C) and right (O) atoms are also shown. The vertical broken lines show the equilibrium bond lengths obtained from previous experiments{\cite{2013HUB}} and from these calculations.}
  \label{pes-CO}
\end{figure*}

\begin{figure}[htbp]
  \centering
  \includegraphics[width=8.6cm]{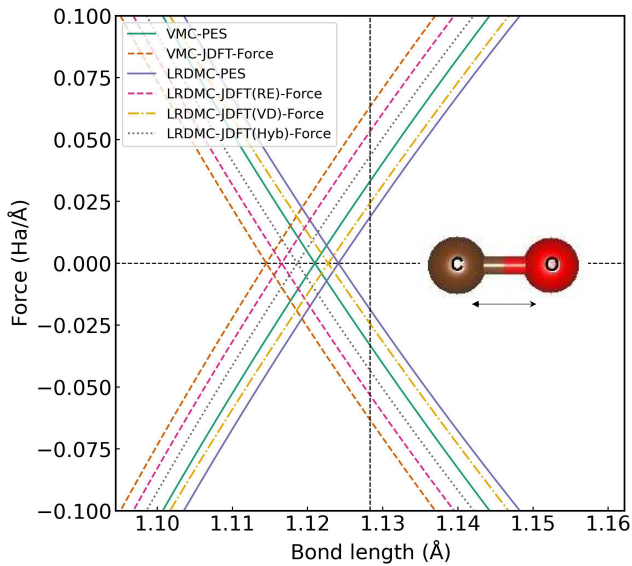}
  \caption{Comparison of the derivatives of the PESs (solid lines) and the atomic forces (broken lines) of the CO dimer obtained from the VMC and LRDMC calculations, corresponding to Fig.~{\ref{pes-CO}}. RE, VD, Hyb refer to the Reynolds{\cite{1986REY}} approximation, the variational-drift{\cite{2014MOR}} approximation, and hybrid forces.{\cite{2003ASS}} The vertical black broken line indicates the equilibrium bond length ($r_{{\rm eq}}$) obtained from the experiment.{\cite{2013HUB}} The CO dimer was depicted by VESTA.{\cite{2011MOM}}}
  \label{force-CO}
\end{figure}

\begin{center}
\begin{table*}[htbp!]
\caption{\label{r_eq_omega_comparison_table}
Equilibrium bond distances $r_{\rm eq}$ (\AA) and harmonic frequencies $\omega$ (cm$^{-1}$) calculated in the present work alongside experimental data.{\cite{2013HUB}} All calculations were carried out using the JDFT ansatz. The subscripts $E$ and $F$ indicate values obtained from the PES and the atomic forces, respectively.}
\scalebox{0.70}{
\begin{tabular}{c|c|cc|cc}
\Hline
Molecule & Method & $r_{{\rm eq}, E}$ (\AA) & $r_{{\rm eq}, F}$ (\AA) & $\omega_{E}$ (cm$^{-1}$) & $\omega_{F}$ (cm$^{-1}$) \\
\Hline
\multirow{6}{*}{H$_2$} 
  &  VMC        & 0.741355(89)  &  0.741322(36)  &  4398.3(5.1)  &  4410.6(1.4)  \\
  &  LRDMC(RE)  & 0.741328(92)  &  0.741319(71)  &  4404.3(6.8)  &  4405.1(2.6)	 \\
  &  LRDMC(VD)  &           -   &  0.741373(72)  &      -        &  4398.9(2.8)	 \\  
  &  LRDMC(Hyb) &           -   &  0.74132(15)   &      -        &  4399.7(5.7)	 \\
\cline{2-6}
  &  Exp.~{\footnotemark[1]} &      0.74144 &            -  &      4401.21  &            - \\  
\Hline
\multirow{6}{*}{LiH} 
  &  VMC        &  1.6031(11)  &  1.60789(28)  &  1384(22)  &  1388.2(2.9)  \\
  &  LRDMC(RE)  &  1.5983(17)  &  1.59936(58)  &  1421(33)  &  1401.8(6.3)  \\
  &  LRDMC(VD)  &          -   &  1.59241(62)  &    -       &  1408.3(6.3)  \\
  &  LRDMC(Hyb) &          -   &  1.5909(12)   &      -     &  1418(12)	\\
\cline{2-6}
  & Exp.~{\footnotemark[1]} &     1.5957  &           -  &    1405.65 &        -     \\ 
\Hline
\multirow{6}{*}{Li$_2$} 
  &      VMC   &  2.7625(41)  &  2.75455(37)  &  335.3(8.3)	 &  333.04(60)  \\
  & LRDMC(RE)  &  2.7135(21)  &  2.73209(40)  &  349.1(5.6)  &  341.18(61)  \\
  & LRDMC(VD)  &           -  &  2.69741(29)  &      -       &  352.18(52)	\\
  & LRDMC(Hyb) &           -  &  2.71075(81)  &      -       &  349.6(1.3)  \\  
\cline{2-6}
  & Exp.~{\footnotemark[1]} &        2.6729 &            - &       351.43 &       -    \\ 
\Hline
\multirow{6}{*}{CO} 
  &      VMC   &  1.12098(39)  &  1.114521(85)  &  2232(13)     &  2277.8(2.0)  \\
  & LRDMC(RE)  &  1.12412(15)  &  1.116556(57)  &  2206.1(4.9)  &  2257.0(1.4)  \\
  & LRDMC(VD)  &         -     &  1.1229(32)    &        -      &  2193(70)     \\
  & LRDMC(Hyb) &          -    &  1.11863(14)   &      -        &  2235.7(3.5)  \\
\cline{2-6}
  & Exp.~{\footnotemark[1]} &     1.128323 &            -  &   2169.81358 &          -  \\ 
\Hline
\multirow{6}{*}{MgO} 
  &       VMC  &  1.7280(26)   &  1.72898(19)  &      830(24)  &  863.4(1.8)  \\
  & LRDMC(RE)  &  1.75334(84)  &  1.72891(16)  &   793.8(4.9)  &  848.6(1.1)  \\
  & LRDMC(VD)  &          -    &  1.7280(32)   &       -       &  896(32)     \\
  & LRDMC(Hyb) &          -    &  1.72884(39)  &       -       &  833.5(3.0)  \\  
\cline{2-6}
  & Exp.~{\footnotemark[1]} &        1.749 &            -  &        785.0 &      -      \\ 
\Hline
\multirow{6}{*}{SiH} 
  &       VMC  &   1.5236(25)   &   1.53281(39)  &   2030(48)  &  2008.9(4.7)  \\
  & LRDMC(RE)  &   1.52303(80)  &   1.52432(36)  &   2061(16)  &  2052.8(4.3)  \\
  & LRDMC(VD)  &       -        &   1.5057(35)   &         -   &  2156(46)     \\
  & LRDMC(Hyb) &          -     &   1.51618(81)  &      -      &  2089.2(9.2)  \\  
\cline{2-6}
  & Exp.~{\footnotemark[1]} &   1.5201     &           -   &      2041.80 &     -        \\ 
\Hline
\end{tabular}
}
\footnotetext[1]{These values are taken from Ref.~\onlinecite{2013HUB}.}
\end{table*}
\end{center}

\begin{figure*}[htbp]
  \centering
  \includegraphics[width=16.4cm]{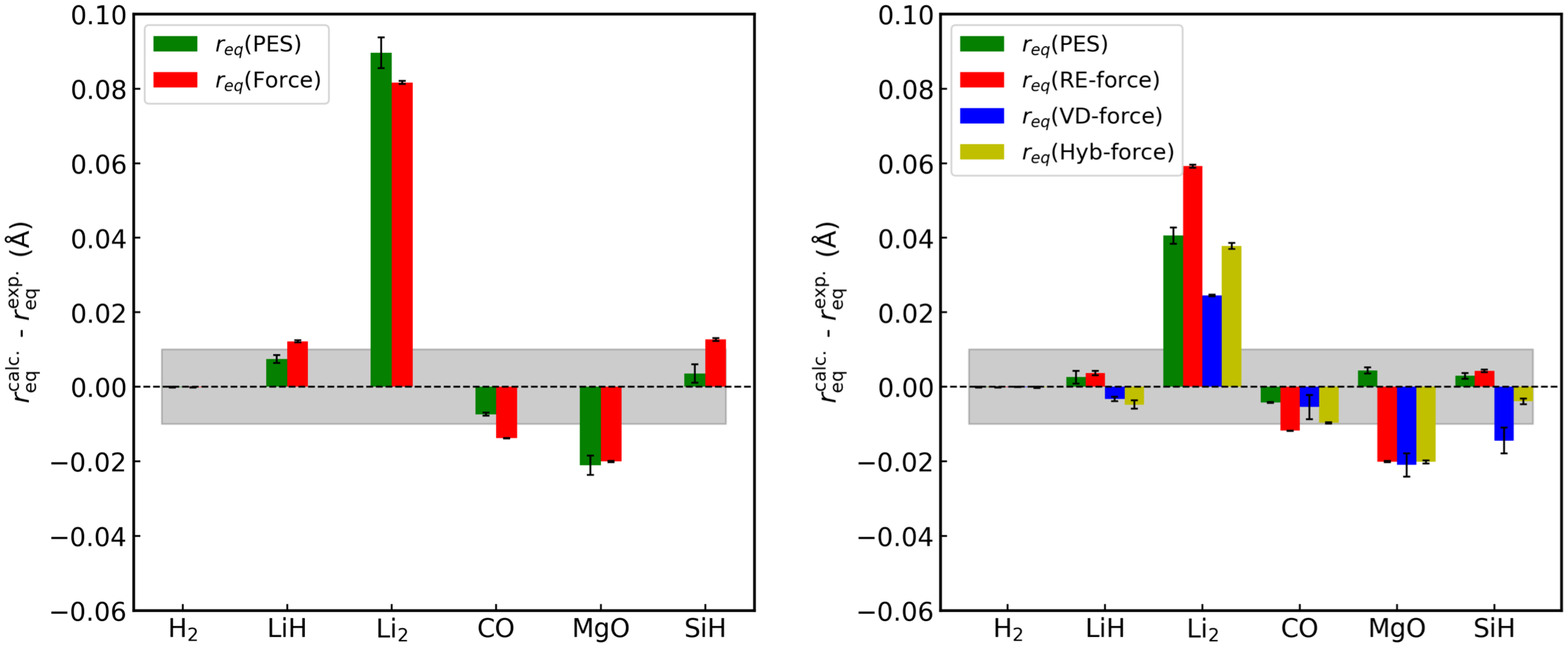}
  \caption{Deviations of the obtained bond lengths $r_{{\rm eq}}$ from the experimentally obtained values.{\cite{2013HUB}} The values obtained from the VMC PESs and atomic forces are shown in the left-hand panel, while those obtained from the LRDMC PESs and atomic forces with the RE, VD, and hybrid forces are shown in the right-hand panel. The corresponding numerical values are shown in Table~{\ref{r_eq_omega_comparison_table}}.}
  \label{r-eq-comparison-fig}
\end{figure*}

\begin{figure*}[htbp]
  \centering
  \includegraphics[width=16.4cm]{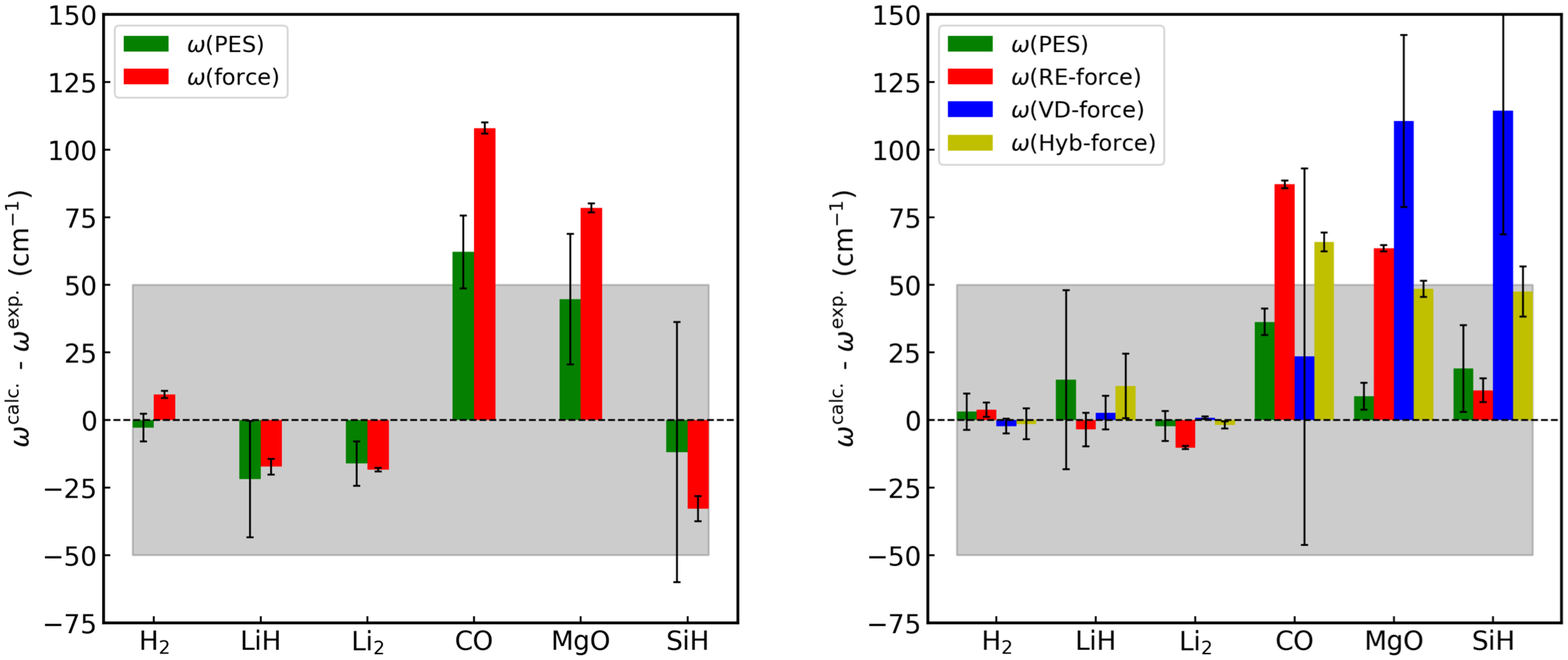}
  \caption{Deviations of the obtained harmonic frequencies $\omega$ from the experimentally obtained values.{\cite{2013HUB}} The values obtained from the VMC PESs and atomic forces are shown in the left-hand panel, while those obtained from the LRDMC PESs and atomic forces with the RE, VD, and hybrid forces are shown in the right-hand panel. The corresponding numerical values are shown in Table~{\ref{r_eq_omega_comparison_table}}.}
  \label{omega-comparison-fig}
\end{figure*}

%
\begin{center}
\begin{table}[hbtp]
\caption{\label{mad}
Mean absolute errors (MAEs) of $r_{{\rm eq}}$ and $\omega$.
}
\begin{tabular}{cc|c|c}
\Hline
$A$ & $B$ & MAE [$r_{{\rm eq}, A} - r_{{\rm eq}, B}$] (\AA) & MAE [$\omega_{A} - \omega_{B}$] (cm$^{-1}$) \\
\Hline
Exp. & $E_{\rm VMC}$ & 0.02150(92) & 27(10) \\
Exp. & $E_{\rm DMC}$ & 0.00913(49) & 14.1(6.4) \\
\Hline
$E_{\rm VMC}$ & $F_{\rm VMC}$ & 0.00490(93) & 20(10) \\
$E_{\rm LRDMC}$ & $F_{\rm VMC}$ & 0.01573(51) & 41.3(6.5) \\
\Hline
$E_{\rm LRDMC}$ & $F_{\rm RE-LRDMC}$ & 0.00882(51) & 23.5(6.6) \\
$E_{\rm LRDMC}$ & $F_{\rm VD-LRDMC}$ & 0.0110(11) & 38(16) \\
$E_{\rm LRDMC}$ & $F_{\rm Hyb-LRDMC}$ & 0.00783(57) & 17.6(7.0) \\
\Hline
\end{tabular}
\end{table}
\end{center}
%

%
%

\section{Results and Discussion}
Figure~{\ref{pes-CO}} shows the PESs of the CO dimer obtained from the VMC calculations and the LRDMC calculations with the RE approximation, wherein the JDFT ansatz was employed. The derivatives of the PESs (i.e., exact forces) and the obtained atomic forces are also shown in Fig.~{\ref{pes-CO}}. The broken vertical lines in the figures indicate the equilibrium bond lengths obtained from the PESs, the atomic forces, and experiments.{\cite{2013HUB}} A comparison of the LRDMC calculations with the VD approximation is shown in the Appendix. Figure~{\ref{force-CO}} focuses on the region near the experimental equilibrium distance. The same comparisons for the other dimers are also shown in the Appendix. We can see at a glance that the left (C) force is exactly opposite to the right (O) force both at the VMC and LRDMC levels. We can readily establish that this is because the SWCT should give the same values of forces with the opposite signs in a dimer as a consequence of the following rigorous and straightforward derivation. The local energy and the logarithm of the wave function, which are explicitly dependent on the atomic and electron positions, satisfy the following translational~symmetries:
\begin{equation}
E_L \left( \cdots, {\bf R}_{\alpha}, \cdots, {\bf r}_{i}, \cdots \right) = E_L \left( \cdots, {\bf R}_{\alpha} + {\bf \Delta}, \cdots, {\bf r}_{i} + {\bf \Delta}, \cdots \right),
\label{trans-el}
\end{equation}
and \begin{equation}
\Psi \left( \cdots, {\bf R}_{\alpha}, \cdots, {\bf r}_{i}, \cdots \right) = \Psi \left( \cdots, {\bf R}_{\alpha} + {\bf \Delta}, \cdots, {\bf r}_{i} + {\bf \Delta}, \cdots \right).
\label{trans-wf}
\end{equation}
They are equivalent, e.g., by differentiating the right-hand sides of Eqs.~({\ref{trans-el}) and (\ref{trans-wf}}) with respect to ${\bf \Delta}$:
\begin{equation}
\sum_{\alpha}{\frac{\partial{E_L}}{\partial{ {\bf R}_{\alpha}}}} + \sum_{i}{\frac{\partial{E_L}}{\partial{ {\bf r}_{i}}}} = 0,
\label{total-deriv-el}
\end{equation}
and
\begin{equation}
\sum_{\alpha}{\frac{\partial{\log \left( \Psi \right)}}{\partial{ {\bf R}_{\alpha}}}} + \sum_{i}{\frac{\partial{\log \left( \Psi \right)}}{\partial{ {\bf r}_{i}}}} = 0.
\label{total-deriv-wf}
\end{equation}
Then, as a consequence of $\sum_{\alpha} {\omega _{\alpha} \left( {{{\mathbf{r}}_i}} \right)} = 1$, the sum of the HF forces in Eq.~(\ref{forces:warp-RE}) over all the atoms in a system is:
\begin{equation}
\braket{\sum_{\alpha} \frac{\partial }{{\partial {{\mathbf{R}}_\alpha}}}{E_L} + \sum_{i,\alpha} {\omega_\alpha\left( {{{\mathbf{r}}_i}} \right)\frac{\partial }{{\partial {{\mathbf{r}}_i}}}{E_L}}
}_{{\Psi_{T}\Phi}} = 0,
\label{HF-part-sum}
\end{equation}
and the sum of the Pulay forces is:
\begin{equation}
2 \braket{ ({{E_L}} - E) \left\{ \sum_{\alpha} \frac{\partial }{{\partial {{\mathbf{R}}_\alpha}}}\log \left( \Psi \right) + \sum_{i,\alpha} \omega_\alpha\left( {{{\mathbf{r}}_i}} \right) {\frac{\partial }{{\partial {{\mathbf{r}}_i}}}\log \left( \Psi \right) } \right\}}_{{\Psi_{T}\Phi}} = 0.
\label{Pulay-part-sum}
\end{equation}
These relations clearly also hold in Eq.~(\ref{forces:warp-VD}). Thus, the total force in a system is always zero regardless of the existence of statistical errors. Indeed, atomic forces in an isolated atom are always zero, and if the SWCT is employed correctly, the right and left forces in a dimer should have the same values with opposite signs, regardless of the presence of statistical errors. We also stress that, since the application of the SWCT does not change the mean values but only reduces variances at the VMC level, the left and right VMC forces should be consistent within their statistical uncertainties even if the SWCT is not applied, as shown in Tables~{\ref{left-right-force-consistent-Li2}} and {\ref{left-right-force-consistent-SiH}}.
%
\begin{center}
\begin{table}[hbtp]
\caption{\label{left-right-force-consistent-Li2}
Left and right VMC forces of the Li$_2$ dimer ($d=2.400$\,\AA) obtained with and without the SWCT.
}
\begin{tabular}{c|c|c}
\Hline
 Li$_2$ & Left force (Ha/\AA) & Right force (Ha/\AA) \\
\Hline
w/SWCT & $-$0.027278(24) & +0.027278(24) \\
wo/SWCT & $-$0.027243(58) & +0.027254(62) \\
\Hline
Absolute difference & 0.000035(63) & 0.000024(66) \\
\Hline
\end{tabular}
\end{table}
\end{center}
%
\begin{center}
\begin{table}[hbtp]
\caption{\label{left-right-force-consistent-SiH}
Left and right VMC forces of the SiH dimer ($d=1.300$\,\AA) obtained with and without the SWCT, where the left and right forces refer to the forces acting on the Si and H atoms, respectively.
}
\begin{tabular}{c|c|c}
\Hline
 SiH & Left force (Ha/\AA) & Right force (Ha/\AA) \\
\Hline
w/SWCT & $-$0.22459(16) & +0.22459(16) \\
wo/SWCT & $-$0.2181(53) & +0.224757(92) \\
\Hline
Absolute difference & 0.0065(53) & 0.00017(19) \\
\Hline
\end{tabular}
\end{table}
\end{center}
%
In Fig.~{\ref{pes-CO}}, the $r_{{\rm eq}}$ value obtained from the LRDMC PES is the closest to the experimental value. This is also true in the other dimers, as will be shown later. Figure~{\ref{pes-CO}} also indicates that self-consistency errors exist both at the VMC and LRDMC levels since the orbitals are not variationally optimized. In the CO dimer, the VD approximation alleviates the self-consistency error more than the RE approximation. This is not true in all six dimers. 
Figures~{\ref{r-eq-comparison-fig}} and {\ref{omega-comparison-fig}} and Table~{\ref{r_eq_omega_comparison_table}} show summaries of the equilibrium distances ($r_{{\rm eq}}$) and harmonic frequencies ($\omega$) estimated from the PESs and forces for the six dimers. The mean absolute errors (MAEs) of $r_{{\rm eq}}$ and $\omega$ are given in Table~{\ref{mad}}. The MAEs of the LRDMC PESs with respect to the experimental values are 0.00913(49)\,\AA\ and 14.1(6.4)\,cm$^{-1}$ for $r_{{\rm eq}}$ and $\omega$, respectively. These are much better than the corresponding VMC values, which are 0.02150(92)\,\AA\ and 27(10)\,cm$^{-1}$ for $r_{{\rm eq}}$ and $\omega$, respectively. This is indeed expected, and it is consistent with the consensus of the QMC community that DMC calculations usually give much better results than VMC calculations.

The self-consistency error is one of the main concerns in QMC force calculations when wave functions are not variationally optimized (i.e., with the JDFT ansatz). The MAEs between a (VMC or LRDMC) PES and the corresponding forces are a measure of the self-consistency error. The MAEs between VMC energies and VMC forces are very small, 0.00490(93)\,\AA\ and 20(10)\,cm$^{-1}$ for $r_{{\rm eq}}$ and $\omega$, respectively, while the MAEs between DMC energies and RE-DMC forces are larger than the VMC ones, 0.00882(51)\,\AA\ and 23.5(6.6)\,cm$^{-1}$ for $r_{{\rm eq}}$ and $\omega$, respectively, implying that the self-consistency error is more severe in the DMC force calculations. The MAEs of the DMC energies and VD-DMC forces are as large as the RE-DMC ones, 0.0110(11)\,\AA\ and 38(16)\,cm$^{-1}$ for $r_{{\rm eq}}$ and $\omega$, respectively. Indeed, the VD forces do not show systematic improvement when compared with the Reynolds forces. The hybrid forces [MAEs: 0.00783(57)\,\AA\ and 17.6(7.0)\,cm$^{-1}$ for $r_{\rm eq}$ and $\omega$] slightly improve the corresponding RE forces [MAEs: 0.00882(51)\,\AA\ and 23.5(6.6)\,cm$^{-1}$ for $r_{\rm eq}$ and $\omega$], but not as drastically as in the original study,{\cite{2003ASS}} which is consistent with the recent DMC-force benchmark test.{\cite{2021TII}}

We see the MAEs of the VD-DMC forces are slightly worse than RE ones overall, but the quality of the VD force depends on each dimer. For instance, as shown in Fig.~{\ref{r-eq-comparison-fig}}, the VD approximation corrects the forces in the right directions for the CO and MgO dimers, but in the cases of LiH, Li$_2$, and SiH, the corrections overshoot. In the original VD paper,{\cite{2014MOR}} it was mentioned that the quality of the trial function crucially affects the bias introduced within the VD approximation. This is why the correction depends on the dimers in this study. We also see that the VD forces show much larger error bars than the RE forces, especially in heavier atoms. For instance, in the SiH dimer at $d = 1.200$\,\AA, the ratio between the variance of force and that of energy ($\sigma_{F}^2/\sigma_{E}^2$) obtained from the same run is $\sim$510 with the VD approximation, while that with the RE approximation is just $\sim$2. The original VD paper also mentioned that the additional Pulay terms \{$\sum_{n-k}^{n-1} \nabla_{\alpha} \cdot [E_L\left({\bf {x}}_{i+1}\right) - E_L\left({\bf {x}}_{i}\right)]$\} increased the variance,{\cite{2014MOR}} but the ratio between the variance of the force and that of the energy calculated in the same run was just $\sim$4$^2 = 16$ at most. This is certainly because pseudopotentials were used in the previous VD work,{\cite{2014MOR}} whereas our calculations are all-electron. Our result implies that the alleviation of the nodal surface divergence by the PW regularization is not sufficient to significantly reduce the error bars due to the presence of the additional Pulay term \{$\sum_{n-k}^{n-1} \nabla_{\alpha} \cdot [E_L\left({\bf {x}}_{i+1}\right) - E_L\left({\bf {x}}_{i}\right)] $\}, especially in all-electron calculations. A new regularization technique that can remove the divergence could alleviate the large-variance problem and make all-electron VD force calculations more practical.

Another concern is whether or not DMC forces are better than VMC forces (i.e., closer to the exact derivative obtained from a DMC PES). If DMC forces do not improve on the VMC ones, it is not worth computing the DMC force with its associated higher computational cost, for instance, to perform structural optimizations or molecular dynamics. The present study shows that the MAEs of VMC and RE-DMC forces with respect to LRDMC energies for $r_{{\rm eq}}$ are 0.01573(51)\,\AA\ and 0.00882(51)\,\AA\, respectively, and those for $\omega$ are 41.3(6.5)\,cm$^{-1}$ and 23.5(6.6)\,cm$^{-1}$, respectively. This implies that RE-DMC forces improve VMC ones, i.e., structural optimizations or molecular dynamics based on RE-DMC forces are better than those based on VMC forces at the JDFT level.

The self-consistency error discussed so far can be removed by optimizing the variational parameters of the wave functions at the VMC level, and this also usually decreases the self-consistency error at the LRDMC level. For example, if a given trial wave function is exact, the RE and VD expressions are consistent with the exact force expression.{\cite{2014MOR}} Table~{\ref{orbopt_r_eq_omega_comparison_table}} shows the $r_{\rm eq}$ and $\omega$ values of the Li$_2$ dimer obtained from the PESs and atomic forces with the JDFT, JSD, and JAGPs ansatz. Comparisons with experimental values are shown in Fig.~{\ref{orbopt_req_omega-comparison-fig}}, and the corresponding PESs are shown in the Appendix (Figs.~{\ref{Li-pes-jdft}}, {\ref{Li-pes-jsd}}, and {\ref{Li-pes-jsagps}}). This result shows that, as expected, the optimized JSD and JAGPs wave functions removed the self-consistency error at the VMC level, and they also alleviated the error at the LRDMC level. Both the VMC and the LRDMC PESs show big discrepancies from the experimental $r_{\rm eq}$ and $\omega$ values with the JDFT ansatz for the Li$_2$ dimer (Fig.~{\ref{orbopt_req_omega-comparison-fig}}), while the optimized JAGPs ansatz almost completely removes these discrepancies. The LRDMC PES with the JAGPs ansatz gives 2.67378(58)\,\AA, 352.4(1.3)\,cm$^{-1}$, and 1.05425(79)\,eV for $r_{\rm eq}$, $\omega$, and the binding energy, respectively. These are in very good agreement with the experimental values of 2.6729\,\AA, 351.43\,cm$^{-1}$, and 1.06(4)\,eV, and also with the CCSD(T) values, which are 2.672\,\AA, 355.10\,cm$^{-1}$, and 1.057\,eV.

\begin{center}
\begin{table*}[hbtp!]
\caption{\label{orbopt_r_eq_omega_comparison_table}
Equilibrium bond distances $r_{\rm eq}$ (\AA) and harmonic frequencies $\omega$ (cm$^{-1}$) of the Li$_2$ dimer obtained with the JDFT, JSD, and JAGPs ansatz. The subscripts $E$ and $F$ indicate values obtained from the PES and the atomic forces, respectively.}
\scalebox{1.00}{
\begin{tabular}{c|c|cc|cc|c}
\Hline
Method & Ansatz & $r_{{\rm eq}, E}$ (\AA) & $r_{{\rm eq}, F}$ (\AA) & $\omega_{E}$ (cm$^{-1}$) & $\omega_{F}$ (cm$^{-1}$) & Binding energy (eV)~{\footnotemark[1]}\\
\Hline
\multirow{3}{*}{VMC} 
  &  JDFT  &  2.7625(41)  &  2.75455(37)   &  335.3(8.3)   &  333.04(60)  & 0.7766(14) \\
  &  JSD   &  2.7548(14)  &  2.75178(14)   &  337.0(2.8)   &  339.18(23)  & 0.79780(90) \\
  &  JAGPs &  2.67941(82)  &  2.67749(15)  &  351.3(1.8)   &  354.28(16)  & 1.03763(66) \\  
\Hline
\multirow{3}{*}{LRDMC-RE} 
  &  JDFT   &  2.7135(21)  &  2.73209(40)   &  349.1(5.6) &  341.18(61) & 0.9759(13) \\
  &  JSD    &  2.7199(15)  &  2.72652(21)   &  345.3(3.8) &  346.16(36) & 0.97258(96) \\
  &  JAGPs  &  2.67378(58) &  2.67435(15)   &  352.4(1.3) &  354.05(21) & 1.05425(79) \\
\Hline
\multirow{3}{*}{LRDMC-VD} 
  &  JDFT   &          -   &  2.69741(29)   &        -  &  352.18(52) & - \\
  &  JSD    &          -   &  2.69394(24)   &        -  &  356.68(38) & - \\
  &  JAGPs  &          -   &  2.67182(19)   &        -  &  353.57(27) & - \\
\Hline
\multirow{1}{*}{CCSD(T)~{\footnotemark[2]}} 
  &  -      &      2.672  &           - &  355.10  &        -     & 1.057 \\
\Hline
\multirow{1}{*}{Exp.} 
  &  -  & 2.6729~{\footnotemark[3]} &  -   & 351.43~{\footnotemark[3]} & - & 1.06(4)~{\footnotemark[4]} \\
\Hline
\end{tabular}}
\footnotetext[1]{Binding energy at the obtained $r_{{\rm eq}, E}$ without the zero-point energy.}
\footnotetext[2]{These values were obtained by {\textsc{Orca}}{\cite{2018NEE}} with extrapolation using cc-pCVnZ basis sets ($n=3,4$).}
\footnotetext[3]{These values are taken from Ref.~\onlinecite{2013HUB}.}
\footnotetext[4]{This value is taken from Ref.~\onlinecite{1989AZI}.}
\end{table*}
\end{center}
%
\begin{figure*}[htbp]
  \centering
  \includegraphics[width=16.4cm]{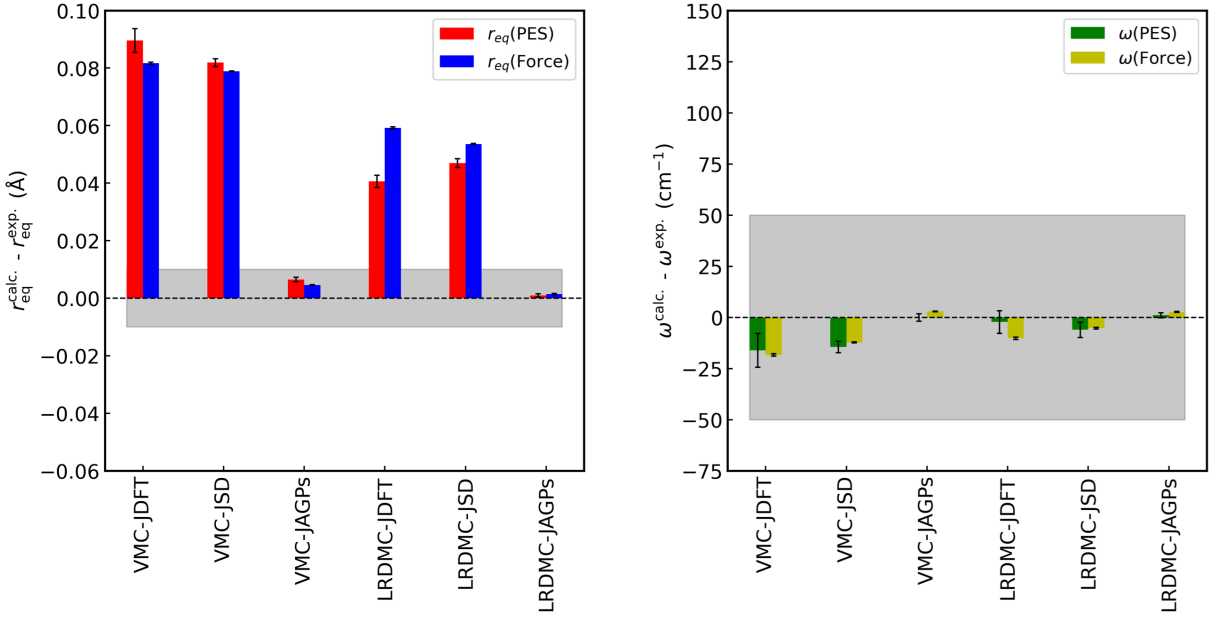}
  \caption{Equilibrium bond lengths (left) and harmonic frequencies (right) of the Li$_2$ dimer obtained with various ansatz and methods. The deviations from the experimental values{\cite{2013HUB}} are plotted. The DMC forces in this figure refer to those with the Reynolds approximation.}
  \label{orbopt_req_omega-comparison-fig}
\end{figure*}

\begin{figure}[htbp]
  \centering
  \includegraphics[width=8.6cm]{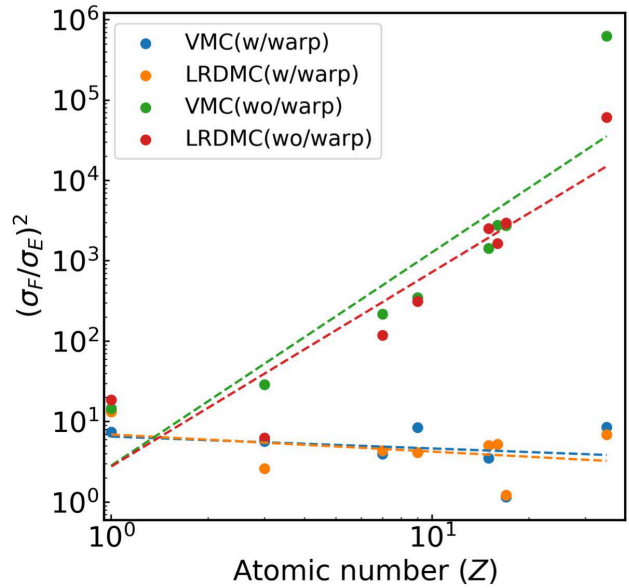}
  \caption{Ratio of variances between force and energy ($\sigma_{F}^2/\sigma_{E}^2$) evaluated with the same run. The VMC and LRDMC forces were evaluated with/without the SWCT, where the LRDMC force refers to the force computed with the RE approximation. The VMC and LRDMC results were fitted with power-law scaling relations.}
  \label{warp-scaling}
\end{figure}
%

As mentioned in the introduction, another important concern in QMC calculations is the scaling of the computational cost with the atomic number ($Z$). A recent VMC and DMC benchmark test carried out by Tiihonen \emph{et~al.}{\cite{2021TII}} suggests that the scaling of the computational cost of forces with respect to the atomic number is much worse than that of energy even if the SWCT is applied.
They claimed the accessible system size scales approximately as $Z^{-2}$.
However, within our implementation, i.e., VMC with the AS regularization and LRDMC with the RE approximation and the PW regularization combined with the SWCT, we did not observe any degradation in the scaling of forces at either the VMC or LRDMC levels for the above all-electron PES calculations.

To study the scalings more systematically, we calculated the energies and forces of homonuclear dimers---H$_2$, Li$_2$, N$_2$, F$_2$, P$_2$, S$_2$, Cl$_2$, and Br$_2$---and computed the ratios between the force and energy variances ($\sigma_{F}^2/\sigma_{E}^2$), wherein $\sigma_{E}$ is the error bar of the total energy of the system and $\sigma_{F}$ is that of the atomic force acting on an atom. The energy and force error bars were evaluated with the same Monte Carlo sampling. The LRDMC forces refer to those with the RE approximation. The variances of energy and force were obtained by applying the reblocking technique to remove the auto-correlations.
Figure~{\ref{warp-scaling}} shows the ratios between the force and energy variances evaluated for each dimer. A summary of the fittings is shown in Table~{\ref{variance_scaling}}. We found that the ratio scales as $Z^{2.65}$ and $Z^{2.42}$ without the SWCT in the VMC and LRDMC calculations, respectively, reproducing the previous report{\cite{2021TII}}.
On the other hand, the ratio is {\it independent} of $Z$ with the SWCT both in the VMC and LRDMC calculations, indicating that the computational cost of QMC forces with respect to $Z$ is no worse than that of energy. Indeed, the accessible system size is not affected by QMC force calculations when the SWCT variance-reduction technique is applied.

The following argument clarifies, in our opinion, why the SWCT is so effective for reducing the variance for large atomic numbers. First of all, the SWCT provides zero forces with zero variances for the calculations of forces of isolated atoms by definition, as already derived. Though the result is trivial in this case, evaluating the zero force without the SWCT becomes problematic in QMC because the derivative of the local energy (required for the HF term) and the derivative of the logarithm of the wave function (required for the Pulay term) scale with $\simeq$$Z$ in the vicinity of nuclei because these functions have length scales of order $1/Z$, inducing large fluctuations of the differentiations in a QMC sampling. Instead, the application of the SWCT is able to suppress the fluctuations because electron coordinations follow the displacement of the atomic position in the vicinity of the nucleus.
This argument clearly holds in complex systems containing several atoms because the damping function ($\omega_{\alpha}$) makes electron coordinations follow an atomic displacement in the vicinity of the nucleus, and the mentioned differentiations are therefore smooth, as in the isolated case. In this way, one can suppress the large fluctuations originating from small electron--ion distances.

\begin{center}
\begin{table}[hbtp]
\caption{\label{variance_scaling}
Scalings of $\sigma_{F}^2/\sigma_{E}^2$ with respect to the atomic number obtained by fitting the VMC and LRDMC forces of the homonuclear dimers (H$_2$, Li$_2$, N$_2$, F$_2$, P$_2$, S$_2$, Cl$_2$, and Br$_2$) with $\alpha Z^{\beta}$.
}
\begin{tabular}{c|c|cc}
\Hline
Method & SWCT & $\alpha$ & $\beta$ \\
\Hline
\multirow{2}{*}{VMC} 
& No & 1.04 & 2.65 \\
& Yes & 1.88 & $-$0.15 \\
\Hline
\multirow{2}{*}{LRDMC} 
& No & 1.15 & 2.32 \\
& Yes & 1.97 & $-$0.24 \\
\Hline
\end{tabular}
\end{table}
\end{center}

\section{Concluding remarks}
In this study, we benchmarked the accuracy of all-electron VMC and LRDMC forces for various mono- and heteronuclear dimers ($1 \le Z \le 35$). Although all our calculations were carried out by considering all the electrons of the corresponding atomic species, we do not expect---nor do we have evidence that---the main conclusions of our work will depend on the presence of pseudopotentials, which are often used in QMC to remove the chemically inert core electrons. The VMC and LRDMC forces were calculated in combination with the SWCT and appropriate regularization techniques to remove the infinite variance problem. The equilibrium distances and the harmonic frequencies of the six dimers (H$_2$, LiH, Li$_2$, CO, MgO, and SiH) obtained from the DMC PESs are very close to the experimental values, but the RE and VD forces are less close as a result of residual self-consistency errors. Self-consistency errors in forces are also seen at the VMC level, but they are much smaller than those in the DMC forces. As shown in the Li$_2$ dimer case, the self-consistency error with the JDFT ansatz can be eliminated by carefully optimizing the determinant part, while until now, there has been no practical way to remove it at the DMC level. Since VMC calculations are sometimes not accurate enough to describe peculiar molecular interactions such as water clusters, developing a practical DMC force calculation without suffering from self-consistency errors is an important challenge for QMC applications. We also found that the ratio of the computational costs between QMC energies and forces scales as $Z^{\sim 2.5}$ without the SWCT, while the application of the SWCT makes this ratio {\it independent} of $Z$. Indeed, the accessible system size is not affected by QMC force calculations when the SWCT variance-reduction technique is employed. This is quite a promising conclusion from the viewpoint of the QMC computation of forces.

\section*{Author declarations}
\subsection*{Conflict of Interest}
The authors have no conflicts to disclose.
\section*{Data availability}
The data that support the findings of this study are available within this article, the supplemental information, and the cited references. \tvb\, is available from S.S.'s website (\href{https://people.sissa.it/~sorella/}{https://people.sissa.it/$\sim$sorella/}) upon request.
\section*{Acknowledgments}
The computations in this work were mainly performed using the supercomputer Fugaku, which was provided by RIKEN through the HPCI System Research Project (Project ID: hp210038).
This work is supported by the European Centre of Excellence in Exascale Computing TREX - Targeting Real Chemical Accuracy at the Exascale. This project has received funding from the European Union's Horizon 2020 research and innovation program (Grant No.~952165).
K.N. and A.R. are grateful for computational resources from the facilities of the Research Center for Advanced Computing Infrastructure at Japan Advanced Institute of Science and Technology (JAIST).
K.N. acknowledges support from the JSPS Overseas Research Fellowships, from Grant-in-Aid for Early Career Scientists (Grant No.~JP21K17752), and from Grant-in-Aid for Scientific Research (Grant No.~JP21K03400).
A.R. gratefully acknowledges the financial support from the MEXT (Monbukagakusho) scholarship.
S.S. acknowledges financial support from PRIN~2017BZPKSZ.
K.N. and S.S. are grateful for valuable comments from C.~Filippi (University of Twente) and S.~Moroni (SISSA) through the TREX collaboration.
The authors are also grateful to A.~Zen (University of Naples Federico~II) for fruitful discussion and for providing us with {\textsc{Orca}} input files for CCSD(T) calculations.

\section*{Appendix}
\makeatletter
\renewcommand{\refname}{}
\renewcommand*{\citenumfont}[1]{#1}
\renewcommand*{\bibnumfmt}[1]{[#1]}
\makeatother

\setcounter{table}{0}
\setcounter{equation}{0}
\setcounter{figure}{0}
\renewcommand{\thetable}{S\Roman{table}}
\renewcommand{\thefigure}{S\arabic{figure}}
\renewcommand{\theequation}{S\arabic{equation}}

\begin{figure*}[htbp]
  \centering
  \includegraphics[width=16.4cm]{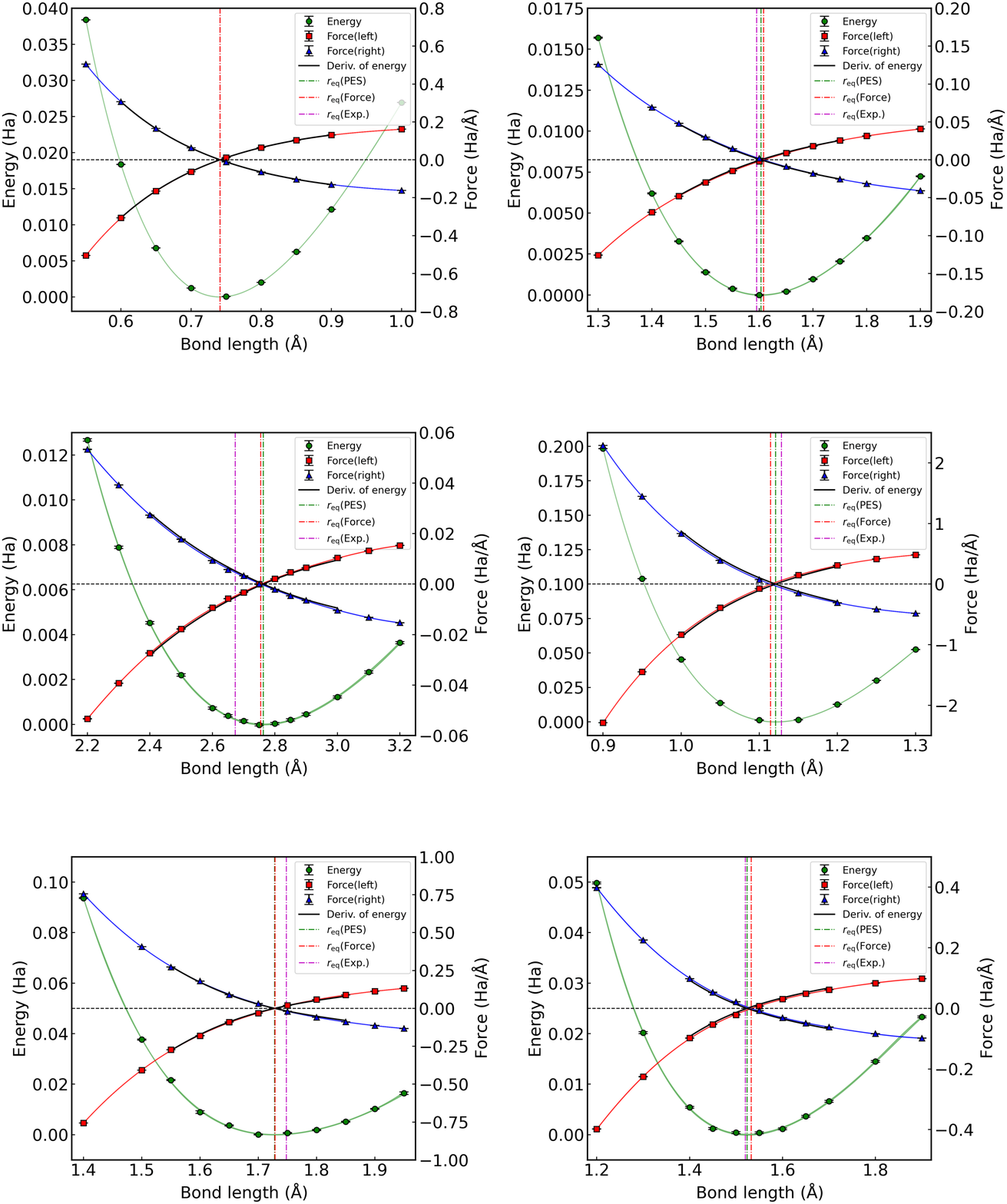}
  \caption{Binding energy curves of the dimers H$_2$ (top left), LiH (top right), Li$_2$ (middle left), CO (middle right), MgO (bottom left), and SiH (bottom right) obtained by VMC(JDFT) calculations.}
  \label{vmc-pes}
\end{figure*}

\begin{figure*}[htbp]
  \centering
  \includegraphics[width=16.4cm]{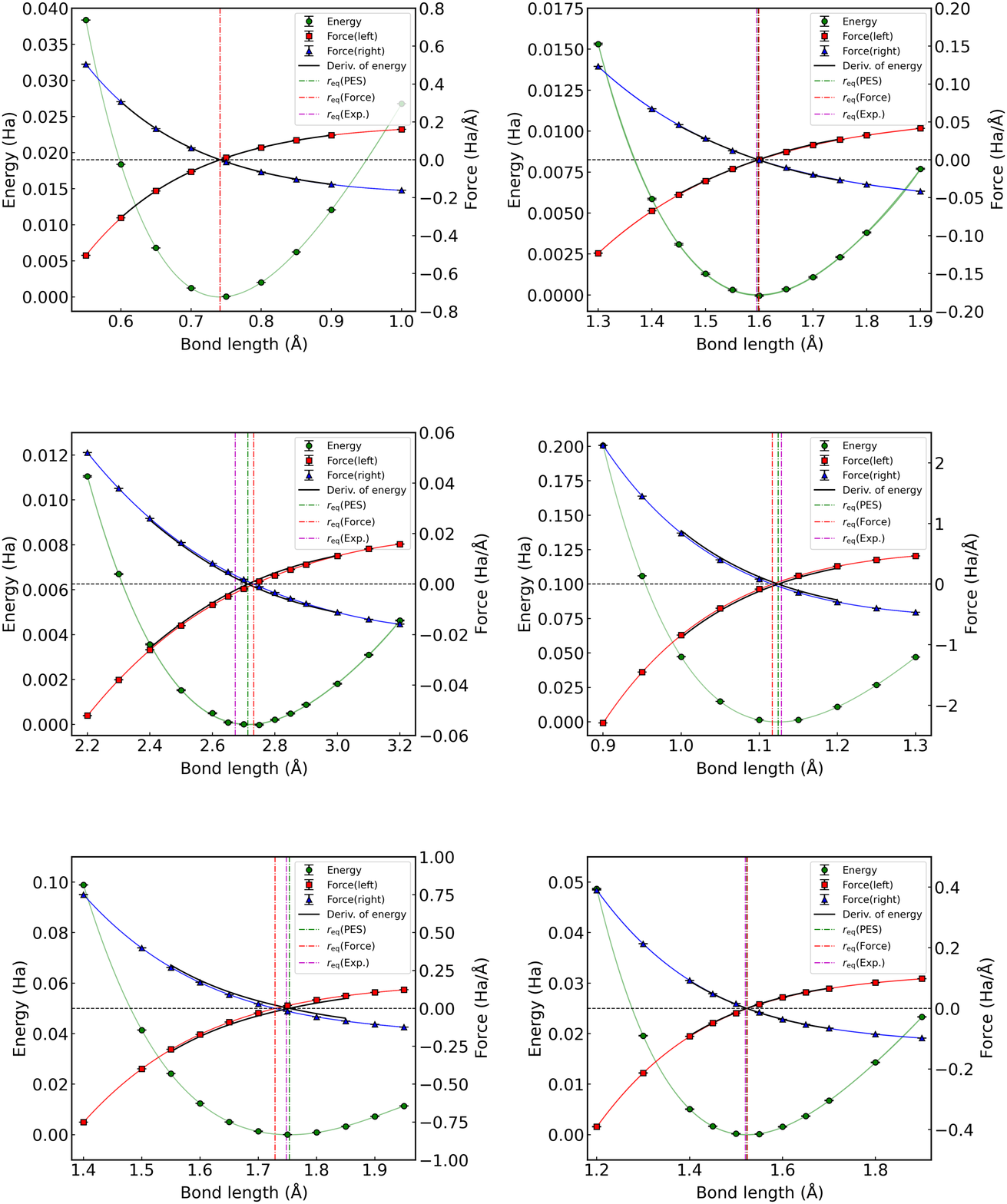}
  \caption{Binding energy curves of the dimers H$_2$ (top left), LiH (top right) Li$_2$ (middle left), CO (middle right), MgO (bottom left), and SiH (bottom right) obtained by LRDMC (JDFT) calculations with the Reynolds approximation.}
  \label{dmc-re-pes}
\end{figure*}

\begin{figure*}[htbp]
  \centering
  \includegraphics[width=16.4cm]{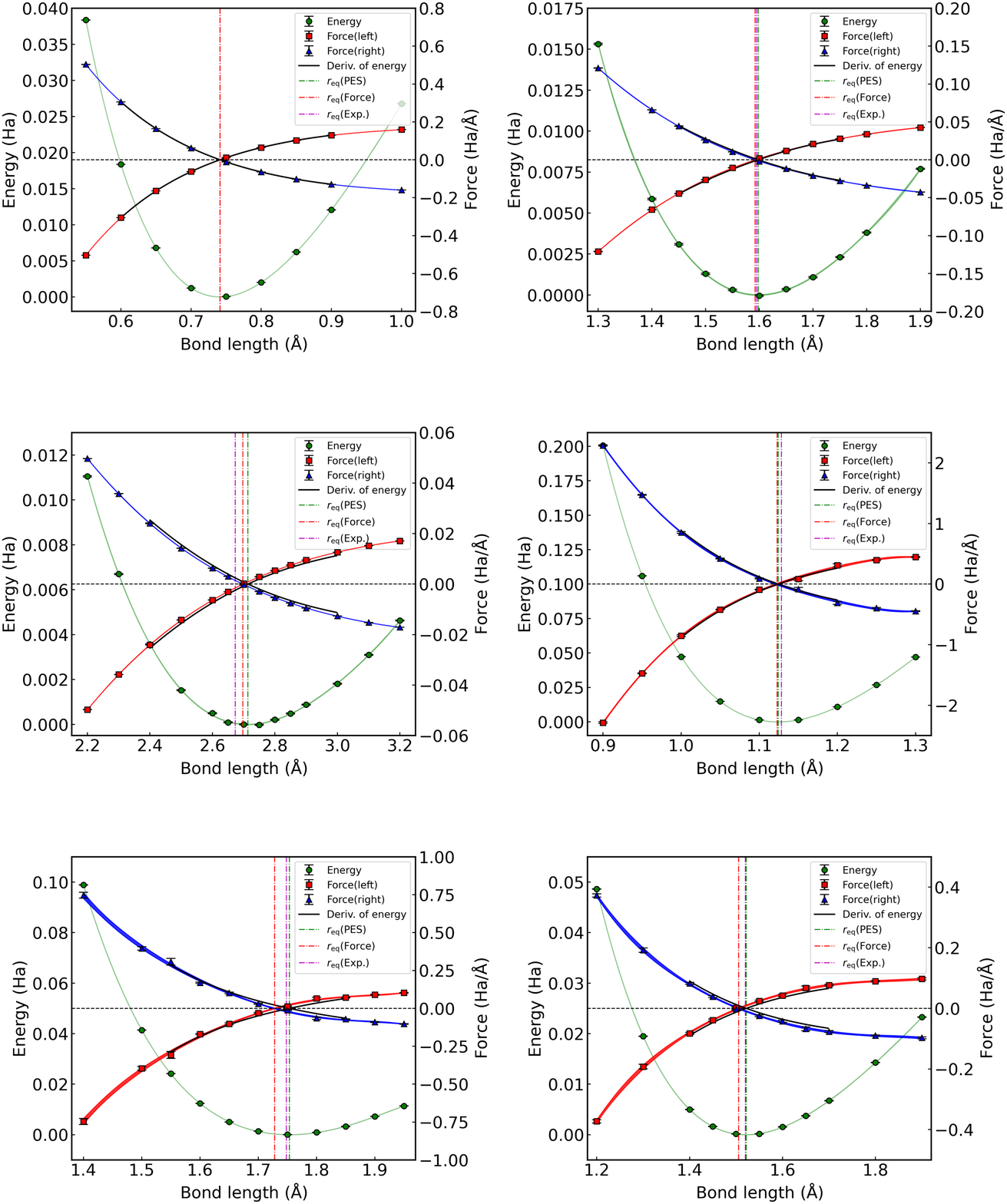}
  \caption{Binding energy curves of the dimers H$_2$ (top left), LiH (top right), Li$_2$ (middle left), CO (middle right), MgO (bottom left), and SiH (bottom right) obtained by LRDMC (JDFT) calculations with the VD approximation.}
  \label{dmc-vd-pes}
\end{figure*}

\begin{figure*}[htbp]
  \centering
  \includegraphics[width=16.4cm]{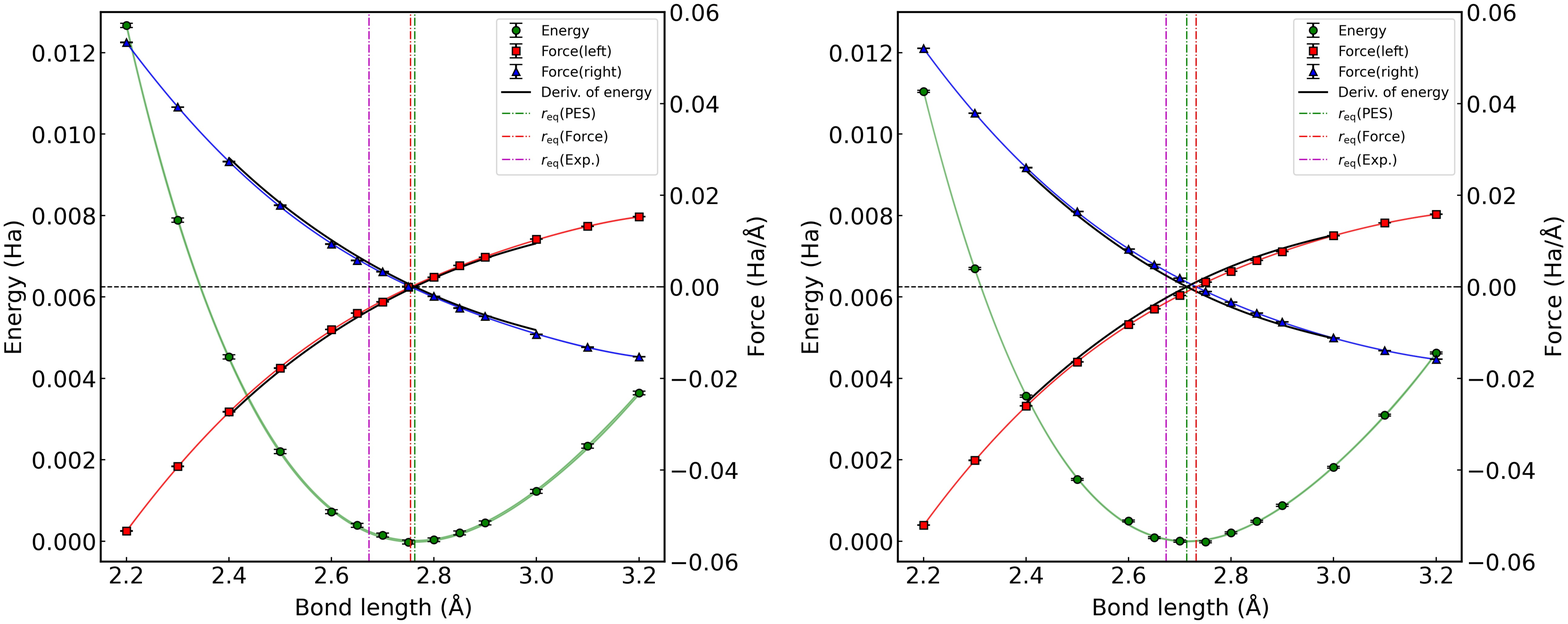}
  \caption{Binding energy curves and their derivatives of the Li$_2$ dimer obtained by VMC (left) and LRDMC-RE (right) calculations with the JDFT ansatz.}
  \label{Li-pes-jdft}
\end{figure*}

\begin{figure*}[htbp]
  \centering
  \includegraphics[width=16.4cm]{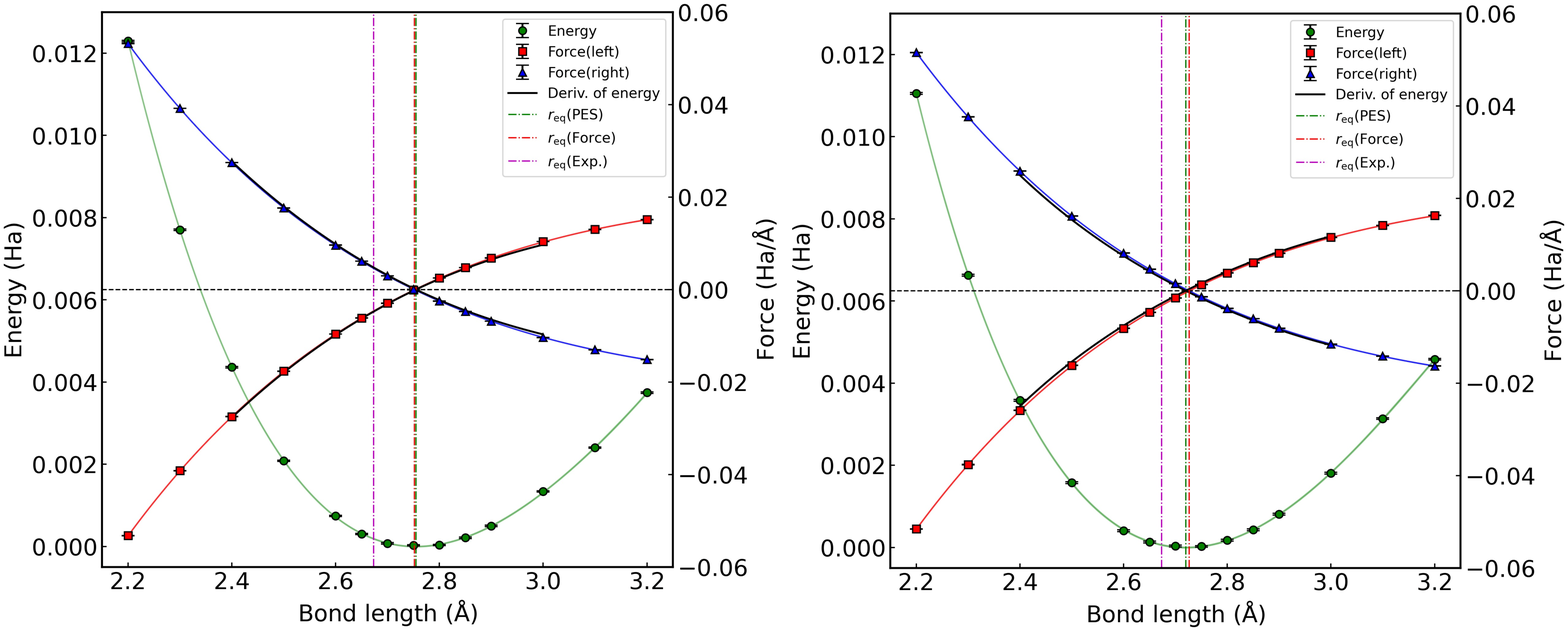}
  \caption{Binding energy curves and their derivatives of the Li$_2$ dimer obtained by VMC (left) and LRDMC-RE (right) calculations with the JSD ansatz.}
  \label{Li-pes-jsd}
\end{figure*}

\begin{figure*}[htbp]
  \centering
  \includegraphics[width=16.4cm]{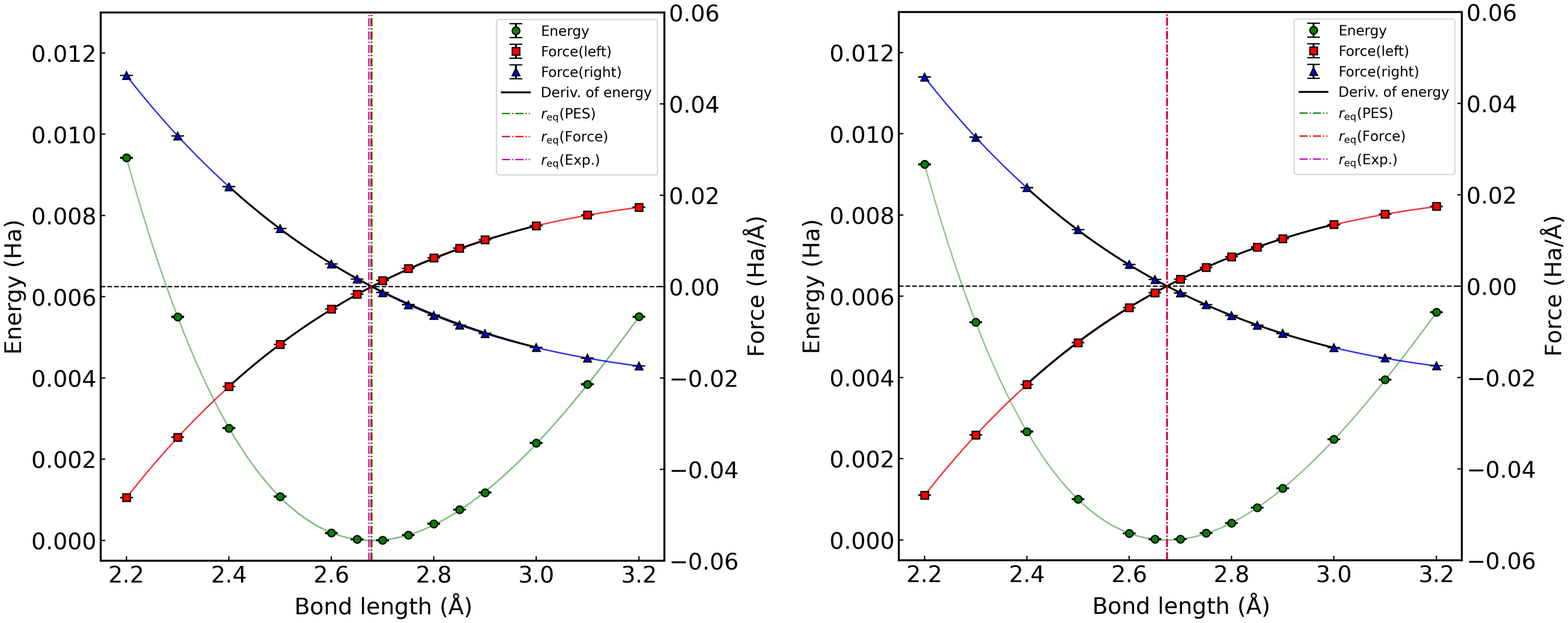}
  \caption{Binding energy curves and their derivatives of the Li$_2$ dimer obtained by VMC (left) and LRDMC-RE (right) calculations with the JAGPs ansatz.}
  \label{Li-pes-jsagps}
\end{figure*}

Figure~{\ref{vmc-pes}} shows the PESs, their derivatives, and the left and right atomic forces obtained by VMC calculations for the six dimers. Figures~{\ref{dmc-re-pes}} and {\ref{dmc-vd-pes}} show those obtained by LRDMC calculations with the RE and VD approximations, respectively. Figures~{\ref{Li-pes-jdft}}, {\ref{Li-pes-jsd}}, and {\ref{Li-pes-jsagps}} are those for the Li$_2$ dimer obtained with various ansatz. We employed sixth- and fifth-order polynomials for fitting the PESs and the atomic forces, respectively. The derivatives of the PESs were obtained from the analytical derivatives of the fitted polynomials. The fittings were performed with the \textsc{polyfit} module implemented in the \textsc{NumPy} package.{\cite{2020HAR}} The equilibrium distances ($r_{{\rm eq}}$) were obtained from the points at which $\frac{dE_{\rm fit}}{dx} =0$ and $F_{\rm fit}=0$, and the harmonic frequencies were obtained from the second derivative of the fitted PES and the first derivative of the fitted forces at the equilibrium distance. The error bars of the equilibrium distances and the harmonic frequencies were estimated by the resampling technique with 3000 independent samplings.
%
%
%
%
%
%
%

\clearpage
\section*{References}
\bibliographystyle{apsrev4-1}
\bibliography{./references}

\end{document}